\documentclass[onecolumn,showpacs,superscriptaddress,twoside,pra]{revtex4}
\usepackage{mathrsfs}
\usepackage{amssymb, amsbsy, amsmath, latexsym, dsfont, array, layout, graphicx,mathrsfs,color}

\newcommand{\one}{\mathds{1}}
\newcommand{\ket}[1]{\left|{#1}\right\rangle}
\newcommand{\bra}[1]{\left\langle{#1}\right|}

\begin{document}

\title{Non-Markovian Decoherent Quantum Walks}
\author{Peng Xue}
\affiliation{Department of Physics, Southeast University, Nanjing
211189, China}
\author{Yongsheng Zhang}
\affiliation{Key Laboratory
of Quantum Information, University of Science and Technology of
China, CAS, Hefei 230026, P. R. China}
\date{\today}

\begin{abstract}
Quantum walk acts obviously different from its classical
counterpart, but decoherence will lessen and close the gap between
them. To understand this process, it is necessary to investigate the
evolution of quantum walk under different situation of decoherence.
In this article, we study a non-Markovian decoherent quantum walk on
a line. In the short time regime, the behavior of the walk deviates
from both idea quantum walks and classical random walks. The
position variance as a measure of quantum walk starts oscillating
from the first several steps and tends to be linear on time and
showing a diffusive spread in the long time limit, which is caused
by the non-Markovian dephasing affecting on quantum correlations
between quantum walker and his coin. We also study both quantum
discord and measurement-induced disturbance as measures of quantum
correlations and observe that both of them oscillate in the short
time regime and tend to be zero in the long time limit. Therefore
quantum walk with non-Markovian decoherence tends to diffusive
spreading behavior in the long time limit, while in the short time
regime it oscillates between a ballistic and diffusive spreading
behavior, and the quantum correlation collapses and revivals due to
the memory effect.
\end{abstract}


\maketitle
\section{Introduction}
The quantum walk (QW) as a generalization of the random walk (RW) is
important in quantum algorithm research~\cite{AAK01,K07} because of
the exponentially speed up the hitting time in glued tree
graphs~\cite{CCD03}. QWs have been demonstrated using nuclear
magnetic resonance~\cite{NMR}, trapped ions~\cite{ion}, single
photons in bulk~\cite{photon} and fibre optics~\cite{S10} and the
scattering of light in waveguide arrays~\cite{B10,S11,P08}. With
decoherence the spreading of the QW is transited from a ballistic to
diffusive scaling~\cite{BCAprl1,BCAprl2} and with quantumness losing
the QW is transited to RW~\cite{XSBL08,XSL09}.

An open quantum system loses its quantumness when information of the
quantum state leaks into the environment around it. The
unidirectional flow of information in which the decoherence and
noise act consistently characterizes a Markovian process~\cite{C93}.
However, there are some systems such as soft- and condensed-matter
systems strongly coupling to the environment and the coupling leads
to a different regime where information also flows back into the
system from the surroundings, which characterizes a non-Markovian
process~\cite{W01}. During a Markovian process, the
distinguishability tends to monotonically decrease for pair of
states, here in this paper, the quantum walker+coin
states~\cite{BCAprl1,BCAprl2,XSBL08,XSL09}. Memory effects caused by
the information flowing back to the system during a non-Markovian
process can temporarily increase it for the walker+coin states. In
this paper, we use quantum correlation between quantum walker and
his coin as a distinguishability to present the behavior of a QW on
a line with a non-Markovian dephasing coin. The quantum correlation
between quantum walker and his coin decays quickly but is
interrupted by revivals. Correspondingly, in the short time regime,
the position variance of a non-Markovian decoherent QW deviates from
both those of the ideal QW and RW. In the long time limit, the
quantum correlation tends to be zero and the position variance shows
a diffusive spread.

This paper is organized as follows: in the next section, we
introduce an idea QW on a line. In Sec.~III, the non-Markovian
dephasing on the coin state---two-level system is shown. We study
the effects from the non-Markovian dephasing on coin. Compared to
both idea QWs and RWs, non-Markovian decoherent QWs show different
behavior in the short time regime. In Sec.~IV, the quantum
correlation between quantum walker and his non-Markovian dephasing
coin is studied via the quantum discord (QD) and measurement-induced
disturbance (MID). It is observed that both of QD and MID show
oscillating behavior in the short time regime and tend to zero in
the long time limit. In Sec.~V, we analyze the behavior of
non-Markovian decoherent QWs in the long time limit and observe that
because of disappearance of the quantum correlation between quantum
walker and his coin, a non-Markovian decoherent QW in the long time
regime shows a diffusive behavior. Finally we summarize the paper
briefly.

\section{An ideal QW on a line}
For an ideal QW on a line~\cite{Kem031,Kem032}, the Hilbert space is
\begin{equation}
\label{eq:H}
    \mathscr{H}=\mathscr{H}_\text{w}\otimes\mathscr{H}_\text{c}
\end{equation}
with the walker Hilbert space~$\mathscr{H}_\text{w}$ spanned by the
position vectors~$\{\ket{x}\}$ and $\mathscr{H}_\text{c}$ the coin
space spanned by two orthogonal vectors which we denote~$\ket{\pm
1}$. Each step by the walker is effected by two subsequent unitary
operators: the coin-flip operator
\begin{equation}
\label{eq:C}
    C=H=\frac{1}{\sqrt{2}}\begin{pmatrix}1&1\\1 & -1\end{pmatrix},
\end{equation}
for~$H$ the Hadamard matrix and the conditional-translation operator
\begin{align}
\label{eq:S}
    F=S\otimes\ket{1}\bra{1}+S^\dagger\otimes\ket{-1}\bra{-1},
\end{align}
where $S\ket{x}=\ket{x+1}$ and $S^\dagger\ket{x}=\ket{x-1}$. The
resultant step operator is $U=F(\one\otimes C)$ with~$\one$ the
identity operator on~$\mathscr{H}_\text{w}$.

The choice of initial state~$\big|\psi(t=0)\big\rangle$ is important
in studies of QWs because the interference features sensitively
depend on the choice of state. This sensitivity is persistent
because the dynamics are unitary hence do not die out. On the other
hand the general properties of QWs do not depend on the choice of
initial state so the choice of initial state is not crucial provided
that the focus is on such characterization.

As we are interested in general properties, the initial state is not
important so we choose the initial state with the walker at the
origin of a line and holding a coin in an equal superposition of
the~$+1$ and~$-1$ states:
\begin{equation}
   \big|\psi(t=0)\big\rangle=\frac{1}{\sqrt{2}}\ket{0}(\ket{1}+i\ket{-1}).
\label{eq:ini}
\end{equation}

After~$t$ steps, the final state of the walker+coin system is
\begin{equation}
    \ket{\psi(t)}=U^t\ket{\psi(0)}.
\end{equation}
We now follow the evolution by performing a Fourier transform of the
evolution operator to the ``momentum''~$k$ space.

The eigenvectors
\begin{equation}
    \ket{k}=\sum_x e^{ikx}\ket{x},
\end{equation}
of~$S$ and $S^\dagger$ in Eq.~(\ref{eq:S}) have the eigenrelations
\begin{equation}
    S\ket{k}=e^{-ik}\ket{k}, S^\dagger\ket{k}=e^{ik}\ket{k}
\end{equation}
for~$k$ a continuous real quantity. The inverse transformation is
\begin{equation}
\ket{x}=\int^\pi_{-\pi}\frac{\text{d}k}{2\pi}e^{-ikx}\ket{k}.
\end{equation}
The walker is initialized at the origin of a line so the walker's
initial state is
\begin{equation}
    \ket{0}=\int^\pi_{-\pi}\frac{\text{d}k}{2\pi}\ket{k}.
\end{equation}
In the~$\{\ket{k}\}$ basis for the walker, the evolution operator
becomes
\begin{equation}
    U\ket{k}\otimes\ket{\Phi}_\text{c}=\ket{k}\otimes U_{k}\ket{\Phi}_\text{c},
\end{equation}
with~$\ket{\Phi}_\text{c}$ the coin state and
\begin{equation}
U_{k}=\begin{pmatrix}
            e^{-ik} & e^{-ik}\\
            e^{ik} & -e^{ik} \\
        \end{pmatrix}.
\end{equation}

The general density operator for the initial state of the system in
the $k$ basis can be expressed as
\begin{equation}
\rho(0)=\int_{-\pi}^\pi\frac{\text{d}k}{2\pi}\int_{-\pi}^\pi\frac{\text{d}k'}{2\pi}
\ket{k}\bra{k'}\otimes\ket{\Phi_0}\bra{\Phi_0}.
\end{equation} The final state after $t$ steps is
\begin{equation}
\rho(t)=\frac{1}{(2\pi)^2}\int_{-\pi}^\pi \text{d}k\int_{-\pi}^\pi
\text{d}k'\ket{k}\bra{k'}\otimes
U_{k}^t\ket{\Phi_0}\bra{\Phi_0}(U_{k}^\dagger)^t.
\end{equation}
In terms of the superoperator
$\mathcal{L}_{kk'}\hat{O}=U_{k}\hat{O}U_{k}^\dagger$,
\begin{equation}
\rho(t)=\frac{1}{(2\pi)^2}\int_{-\pi}^\pi \text{d}k\int_{-\pi}^\pi
\text{d}k'\ket{k}\bra{k'}\otimes\mathcal{L}^t_{kk'}\ket{\Phi_0}\bra{\Phi_0}.
\end{equation}

The walker's position is on a line labeled~$x$ with the initial
position localized at~$0$. Measurement of the walker's position
corresponds to the projection-valued measure~$\{|x\rangle\langle
x|\otimes\one;x\in\mathbb{Z}\}$ on the walker reduced state (after
tracing out the coin).

The probability $P(x;t)$ that the walker will be found at the position~$x$ is
\begin{equation}
    P(x;t)=\text{Tr}\left\{\ket{x}\bra{x}\otimes \one_\text{c}\rho(t)\right\},
\end{equation}
which is obtained by tracing over the coin of the walker+coin state
and then measuring the walker's position. We can characterize
$P(x;t)$ by the moments of this position distribution~$\langle
x^m\rangle$. The mean~$\langle x\rangle$ and variance
\begin{equation}
\text{var}=\langle x^2\rangle-\langle x\rangle^2
\end{equation} can be used as the measure of QWs and show the signature of QWs
compared to RWs. Instead of variance we can also use its square
root, namely dispersion
\begin{equation}
\label{eq:dispersion}
    \sigma=\sqrt{\text{var}}=\sqrt{\langle x^2\rangle-\langle x\rangle ^2}.
\end{equation}
For a RW,~$\sigma\sim \sqrt{t}$, which is characteristic of
diffusive motion, whereas, for a QW, a quadratic enhancement is
achieved:~$\sigma\sim t$~\cite{Kem031,Kem032,BCAprl1,BCAprl2}.

\section{A QW on a line with a Non-Markovian Dephasing Coin in the short time regime}

\begin{figure}
   \includegraphics[width=9cm]{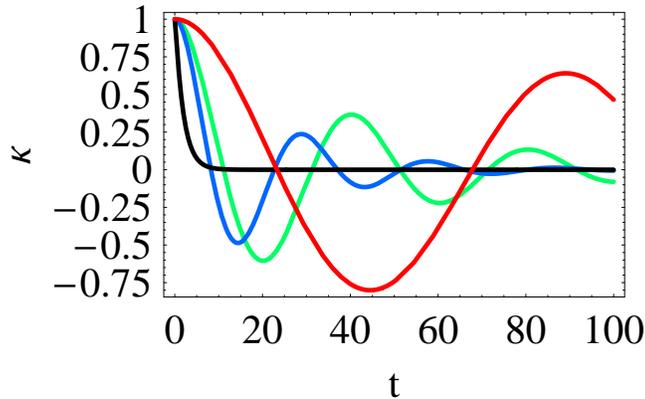}
   \caption{(Color online) The rate $\kappa$ as a function of time $t$
   with $g_0=1$ and different $\eta$: $\eta=0.01$ in red, $\eta=0.05$ in green,
   $\eta=0.1$ in blue presenting non-Markovian
   dephasing and $\eta=10$ in black presenting Markovian dephasing.}
   \label{kappa}
\end{figure}

We now generalize to allow for decoherence. Open quantum system
loses its quantumness when information about the state leaks into
its surroundings. Unidirectional flow of information characterizes a
Markovian process~\cite{C93}. Whereas, for a non-Markovian
process~\cite{CCD03}, information also flows back into the system.
Recently, though it is not well understood yet, non-Markovian
processes with memory have become of central importance in the study
of open system for both theory~\cite{WECC08,RHP10,LPB10,BLP09} and
experiments~\cite{exp1,exp2,Xu10,Xu,Tang,Liu11}.

Suppose that before each unitary flip of the coin, a completely
positive map is performed on the coin. We consider a pure dephasing
quantum process for which the density matrix
$\rho_\text{c}(t)=\text{Tr}_\text{w}\rho(t)$ of the open system
evolves according to the master equation
\begin{equation}
\frac{\text{d}}{\text{d}t}\rho_\text{c}(t)=
-i\frac{\epsilon(t)}{2}\left[\sigma_z,\rho_\text{c}(t)\right]
+\frac{\gamma(t)}{2}\left[\sigma_z\rho_\text{c}(t)\sigma_z,\rho_\text{c}(t)\right].
\label{eq:mastereq}
\end{equation}
Here $\epsilon(t)$ represents the time-dependent energy shift and
$\gamma(t)$ the time-dependent rate of the decay channel described
by the Pauli operator $\sigma_z$. The dephasing process influences
the coherence between the walker and his coin and the evolution can
be described by the time-dependent function
\begin{equation}
\kappa(t)=\exp\left\{-\int_0^t\text{d}t'\left[\gamma(t')+i\epsilon(t')\right]\right\}
\end{equation}
which is connected to the energy shift and the decay rate of the
master equation~(\ref{eq:mastereq}) by the relations
\begin{equation}
\epsilon(t)=-\text{Im}\left[\frac{\dot{\kappa}(t)}{\kappa(t)}\right],
\gamma(t)=-\text{Re}\left[\frac{\dot{\kappa}(t)}{\kappa(t)}\right].
\end{equation}

If the relaxation rates are positive function $\gamma(t)\geq 0$, the
generator in Eq.~(\ref{eq:mastereq}) is in Lindblad form for each
fixed $t\geq0$. Such a process with $\gamma(t)\geq0$ may be called
time-dependent Markovian. The generalized master equation involving
a certain memory kernel can describe a non-Markovian process, in
which the rate $\gamma(t)$ must take on negative values for some
interval of time. In the Markovian regime, information of the coin
state leaks into its surroundings and $|\kappa(t)|$ is a
monotonically decreasing function of time. In the non-Markovian
environment, in contrast, information also flows back into the
system of coin state and a revival of the distinguishability can be
observed in the time evolution. With the definition of
non-Markovianity, we see that an increase of $|\kappa(t)|$ leads to
a negative rate $\gamma(t)$ in the generator of the master
equation~(\ref{eq:mastereq}). As an example, we consider the case of
a Lorentzian reservoir spectral density which is on resonance with
the coin qubit transition frequency and leads to an exponential two
point correlation function $f\left(t\right)=g_0\eta e^{-\eta|t|}/2$,
where $g_0$ describes the coupling strength between the coin qubit
and environment and $\eta$ denotes the spectral width (here we
consider the coin qubit coupled to a damped Jaynes-Cummings model).
The rate $\kappa\left(t\right)$ is defined as the solution of the
integrodifferential equation,
\begin{equation}
\frac{\text{d}}{\text{d}t}\kappa\left(t\right)=-\int_{0}^t\text{d}t'f\left(t-t'\right)\kappa\left(t'\right)
\end{equation}
corresponding to an initial condition $\kappa\left(0\right)=1$. We
obtain
\begin{equation}
\kappa\left(t\right)=e^{-\eta t/2}\left[\cosh\left(\frac{d
t}{2}\right)+\frac{\eta}{d}\sinh\left(\frac{d t}{2}\right)\right],
\label{eq:kappa}
\end{equation}
where $d=\sqrt{\eta^2-2g_0\eta}$. We see that for weak couplings,
$g_0<\eta/2$, $|\kappa\left(t\right)|$ decreases monotonically. The
correlation function $f(t)$ tends to $\delta(t)$ with $\eta$
increasing. For $\eta\rightarrow \infty$, $f(t)=\delta(t)$ and the
noise of different time is not correlated. Whereas, in the strong
coupling regime $g_0>\eta/2$, $d$ turns to be a complex number and
$|\kappa\left(t\right)|$ can be rewritten as
\begin{equation}
\kappa\left(t\right)=e^{-\eta t/2}\left[\cos\left(\frac{d'
t}{2}\right)+\frac{\eta}{d'}\sin\left(\frac{d't}{2}\right)\right]
\end{equation} with $d'=\mathcal{I}m\left\{d\right\}$ and
starts to oscillate, showing non-Markovian
behavior~\cite{WECC08,RHP10,LPB10,BLP09,exp1,exp2,Xu10,Xu,Tang,Liu11}.

The plots of $\kappa(t)$ are shown in Fig.~\ref{kappa}. In the
Markovian regime, for example $g_0=1$ and $\eta=10$, the flow of
information goes only from the system into the environment, and
$\kappa(t)$ decreases monotonically. In the non-Markovian regime,
such as $g_0=1$, and $\eta=0.01,\text{ }0.05,\text{ }0.1$
respectively, memory effects however can temporarily increase it.
The function $\kappa(t)$ decreases faster at $t=0$ but will be
interrupted by revivals which are due to memory effects on quantum
correlations between the quantum walker and his coin.

The corresponding dynamical map which maps the coin state
$\rho_\text{c}(0)$ to the state $\rho_\text{c}(t)$ at time $t$
\begin{equation}
\rho_\text{c}^{00}=\rho_\text{c}^{00},\rho_\text{c}^{01}=\kappa^*(t)\rho_\text{c}^{01},
\rho_\text{c}^{10}=\kappa(t)\rho_\text{c}^{10},\rho_\text{c}^{11}=\rho_\text{c}^{11}.
\end{equation}
This map can also be written by a set of operators
$\left\{A_1,A_2\right\}$ on the coin degree of freedom which satisfy
\begin{equation}
\sum_{n=1,2}A^\dagger_n A_n=\one,
\end{equation}
where
\begin{align}
& A_1=\frac{\left(1-\sqrt{1-|\kappa(t)|^2}\right)|\kappa(t)|}
{\kappa^*(t)\sqrt{|\kappa(t)|^2+\left(1-\sqrt{1-|\kappa(t)|^2}\right)^2}}\ket{0}\bra{0}
+\frac{|\kappa(t)|}{\sqrt{|\kappa(t)|^2+\left(1-\sqrt{1-|\kappa(t)|^2}\right)}}\ket{1}\bra{1}, \nonumber\\
&A_2=\frac{|\kappa(t)|}{\sqrt{|\kappa(t)|^2+\left(1-\sqrt{1-|\kappa(t)|^2}\right)}}\ket{0}\bra{0}
+\frac{\left(1-\sqrt{1-|\kappa(t)|^2}\right) |\kappa(t)|}
{\kappa(t)\sqrt{|\kappa(t)|^2+\left(1-\sqrt{1-|\kappa(t)|^2}\right)^2}}\ket{1}\bra{1}.
\end{align}
Considering the initial state is
\begin{equation}
\rho(0)=\int_{-\pi}^\pi\frac{\text{d}k}{2\pi}\int_{-\pi}^{\pi}\frac{\text{d}k'}{2\pi}
\ket{k}\bra{k'}\otimes\ket{\Phi_0}\bra{\Phi_0}.
\end{equation}
Let the QW proceed for $t$ steps. Then the state evolves to
\begin{align}
\rho\left(t\right)&=\int_{-\pi}^\pi\frac{\text{d}k}{2\pi}\int_{-\pi}^{\pi}\frac{\text{d}k'}{2\pi}
\ket{k}\bra{k'}\otimes\sum_{n_1,...,n_t}U_k
A_{n_t}\cdot\cdot\cdot\cdot U_k
A_{n_1}\ket{\Phi_0}\bra{\Phi_0}A^\dagger_{n_1}U^\dagger_k\cdot\cdot\cdot
A^\dagger_{n_t}U^\dagger_k \nonumber\\
&=\int_{-\pi}^\pi\frac{\text{d}k}{2\pi}\int_{-\pi}^{\pi}\frac{\text{d}k'}{2\pi}
\ket{k}\bra{k'}\otimes\mathcal{L}_{kk'}^t\ket{\Phi_0}\bra{\Phi_0}.
\end{align}
Note that for $k=k'$ this superoperator preserves the trace. This
implies that
\begin{equation}
\text{Tr}\left(\mathcal{L}_{kk}^t\hat{O}\right)=\text{Tr}\left(\hat{O}\right)
\end{equation}
for any operator $\hat{O}$.

\begin{figure}
   \includegraphics[width=9cm]{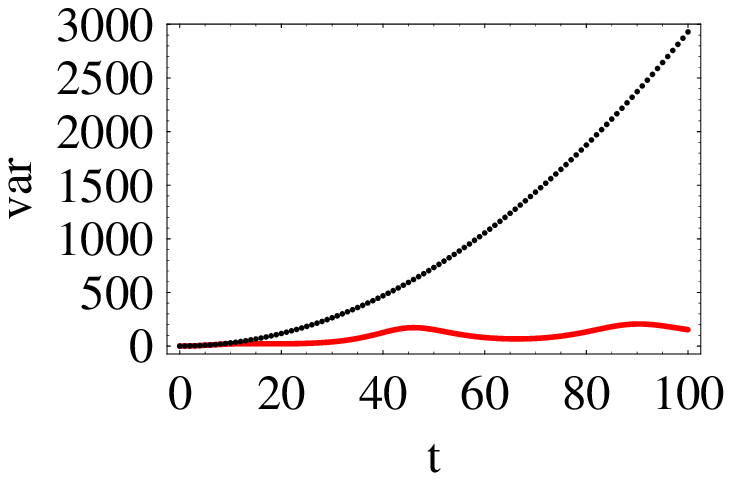}
   \includegraphics[width=9cm]{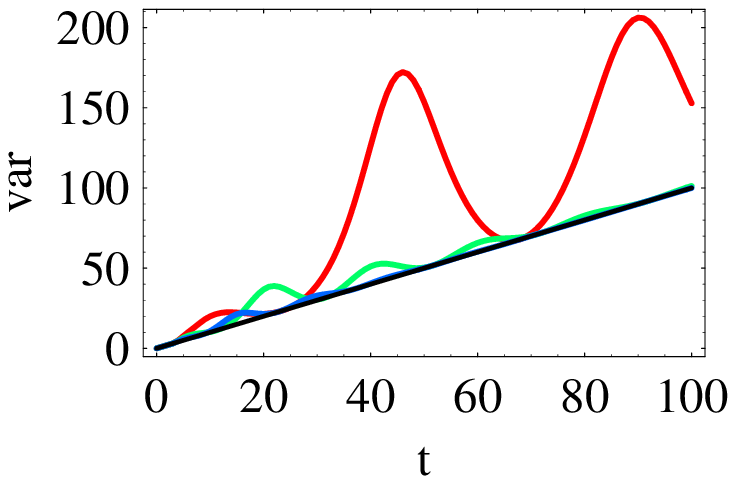}
   \caption{(Color online) The variances for a QW on a line after~$t=100$ steps with
   a perfect coin in dotted-line, with a non-Markovian dephasing coin
   with parameters $g_0=1$ and $\eta=0.01$ in
   red and $\eta=0.05$ in green, $\eta=0.1$ in blue and $\eta=10$ in black.
   Here we choose $\left(\ket{0}+i\ket{1}\right)/\sqrt{2}$ as an initial coin state for calculations.
   (a) Comparison of variances between the RW and QW with a non-Markovian dephasing coin
   with parameters $g_0=1$ and $\eta=0.01$. (b) Comparison of variances between the QW with a non-Markovian dephasing coin
   with $g_0=1$ and $\eta=0.01$, $\eta=0.05$, $\eta=0.1$ and $\eta=10$.}
   \label{var}
\end{figure}

Now we consider the effect of the non-Markovian dephasing on coin.
As we know, without decoherence the variance of position
distribution of QW is quadratically dependent on time $t$. With
Markovian dephasing on coin, in the long time limit the transition
from a ballistic to a diffusion spreading behavior is observed. In
this paper, we will show the behavior of a QW with non-Markovian
dephasing coin. The variance is used as a signature of the QW
compared to RW. In the short time regime we can calculate the
variance of position distribution numerically. In Fig.~\ref{var}
from the numerical results we can see the variance as a function of
time starts to oscillate from the first several steps and the period
of oscillations depends on the coefficient $d$. Compared to that of
ideal QW, in the short time regime ($t\lesssim 100$) the variance of
the QW with non-Markovian dephasing coin deviates from a ballistic
spreading behavior. However, in comparison with RW, the oscillating
variance neither shows a diffusion behavior, which is due to effect
on quantum correlations between quantum walker and his coin from the
non-Markovian decoherence. From the numerical calculations the
position variance of a RW distribution is the lower bound of that of
QW with non-Markovian dephasing coin.

\begin{figure}
   \includegraphics[width=8.5cm]{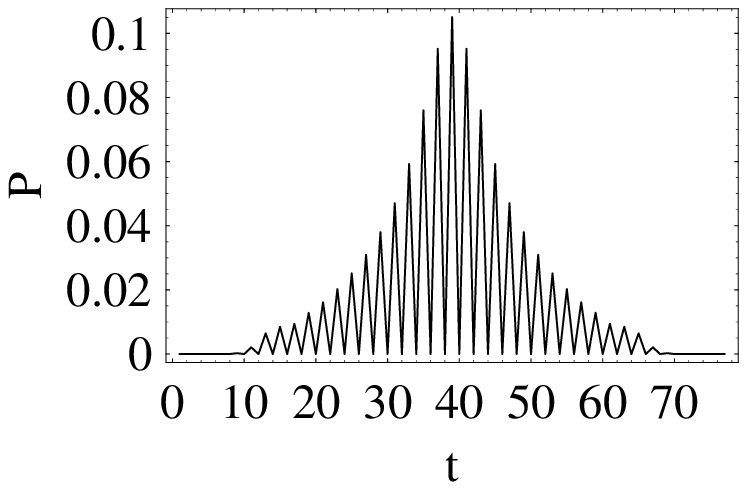}
   \includegraphics[width=8.5cm]{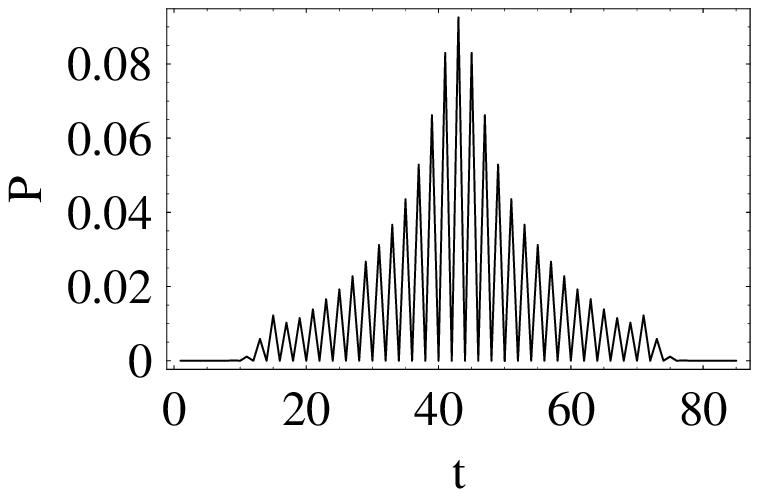}
   \includegraphics[width=8.5cm]{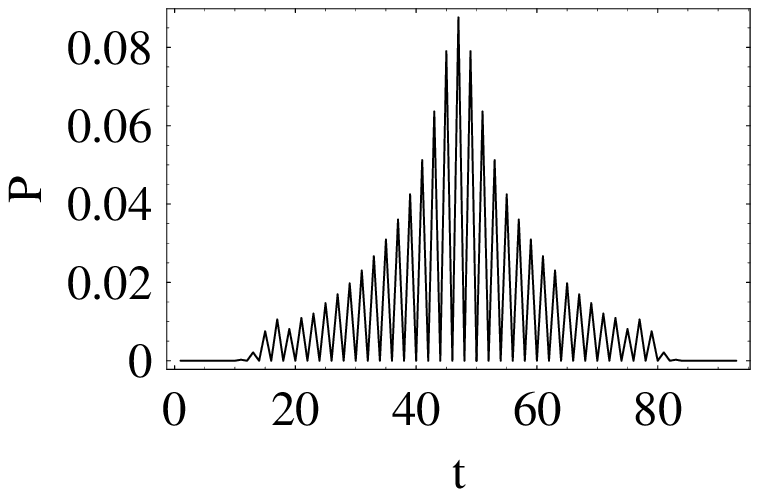}
   \includegraphics[width=8.5cm]{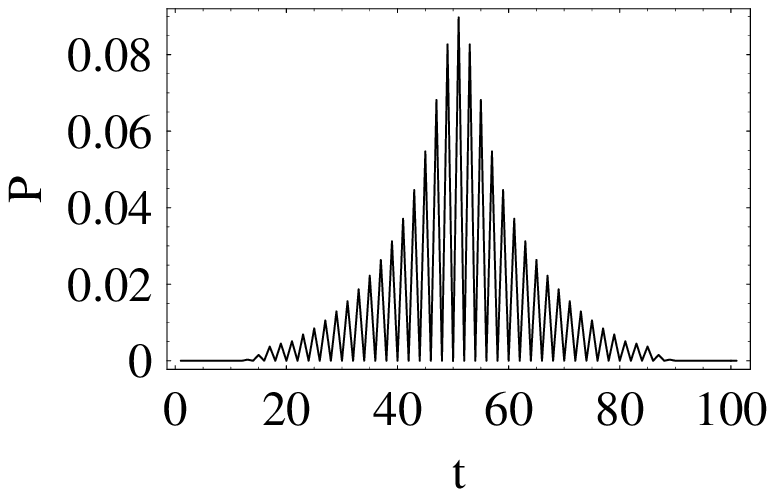}
   \caption{(Color online) The position distribution for a QW on a line with a non-Markovian dephasing coin
   with parameters $g_0=1$ and $\eta=0.01$ at the 38th step (a), the
   42nd step (b), the 46th step (c) and the 50th step (d).}
   \label{position}
\end{figure}

In Fig.~\ref{position}, we plot position distribution for a QW with
a non-Markovian dephasing coin ($g_0=1$ and $\eta=0.01$) at the
38th, 42nd, 46th, and 50th steps. At the 38th step, the position
distribution shows RW behavior, and at the 42nd and 46th steps,
there are two small peaks higher than the others nearby in the
position distribution. It shows the memory effect from non-Markovian
decoherence draws QW behavior back a little. The small peaks
disappear at the 50th step. The plots of position distribution show
the behavior of QW with non-Markovian dephasing coin oscillates
between the quantum (ballistic) and classical (diffusive) behaviors.

\section{Quantum correlations dynamics in non-Markovian environment}

The signature of a decoherent discrete-time QW on a line can be
demonstrated in various ways such as the position probability
distribution becoming increasingly Gasussian, with a concomitant
fall in the standard deviation. Both of them do not recover the
correlations between the quantum walker and his coin, which is known
as the reason causing the quantum behavior of QWs. In this section
we study the quantum correlation of a decoherent QW by two popular
measures, measurement-induced disturbance (MID)~\cite{L1,L2,L3} and
quantum discord (QD)~\cite{OZ01,HV01}.

Given the walker+coin state $\rho(t)$, let the density matrix for
the system be diagonalized to
\begin{equation}
    \rho_i=\sum_j p_i^j\Pi_i^j,
\end{equation}
for $i=\text{w},\text{c}$, where $\{\Pi_i^j\}$ is a complete
projection-valued measure (i.e.\ using von Neumann measurements) for
walker~w or coin c. Summing over joint projections on the state
yields the diagonalized state
\begin{equation}
    \Pi\rho(t)=\sum_{k}\one_\text{w}\otimes\Pi_\text{c}^k\rho(t)\one_\text{w}\otimes\Pi_\text{c}^k.
\end{equation}
The leftmost $\Pi$ is an operator on the density matrix that
diagonalizes it in the spectral basis corresponding to $\Pi_i^j$
projective measurements. The operator~$\Pi$ can also be described as
a `local measurement strategy'.

Correlations between the walker $\rho_\text{w}$ and coin state
$\rho_\text{c}$ are regarded as classical if there is a unique local
measurement strategy $\Pi$ leaving $\Pi\rho(t)$ unaltered from the
original walker+coin state~$\rho(t)$~\cite{L1,L2,L3}. We ascertain
whether the walker+coin state is `quantum' by determining whether a
local measurement strategy exists that leaves the state unchanged.

The degree of quantumness is given by the MID~\cite{L1,L2,L3}
\begin{equation}
\label{eq:MID}
    Q\left[\rho(t)\right]=I\left[\rho(t)\right]-I\left[\Pi\rho(t)\right],
\end{equation}
for
$I\left[\rho(t)\right]=S\left[\rho_\text{w}(t)\right]+S\left[\rho_\text{c}(t)\right]-\left[\rho(t)\right]$
the quantum mutual information and
$S\left[\rho(t)\right]=-\text{Tr}\left[\rho(t)\log_2\rho(t)\right]$
denotes von Neumann entropy. By construction~$\Pi\rho(t)$ is
classical. Thus, the MID is the difference between the quantum and
classical mutual information, which quantify total and classical
correlations. Accordingly, Eq.~(\ref{eq:MID}) is interpreted as the
difference between the total and classical correlations, which are
represented by the quantum mutual information and the mutual
information.

QD~\cite{OZ01,HV01} is defined by
\begin{equation}
D\left[\rho(t)\right]=I\left[\rho(t)\right]-\sup_{\Pi}I\left[\Pi\rho(t)\right].
\end{equation}
The MID has an operational definition, unlike QD, it also tends to
overestimate non-classicality because of lack of optimization over
local measurements. Applied to QWs, we find that MID, while acting
as a loose upper bound on QD, still tends to reflect well trends in
the behavior of the latter.

\begin{figure}
   \includegraphics[width=9cm]{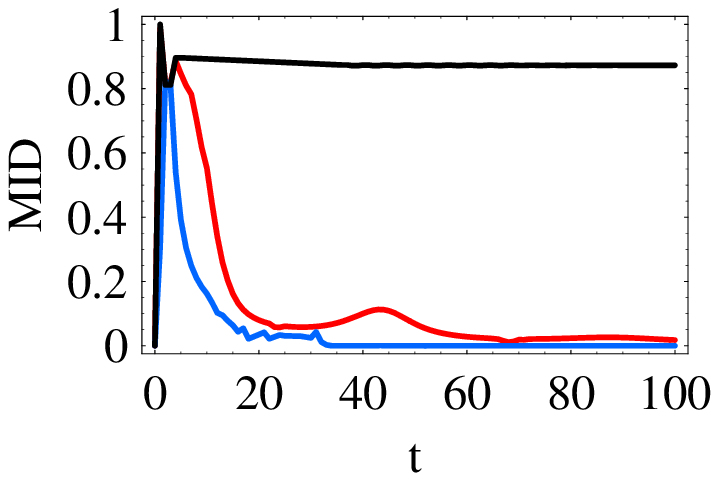}
   \includegraphics[width=9cm]{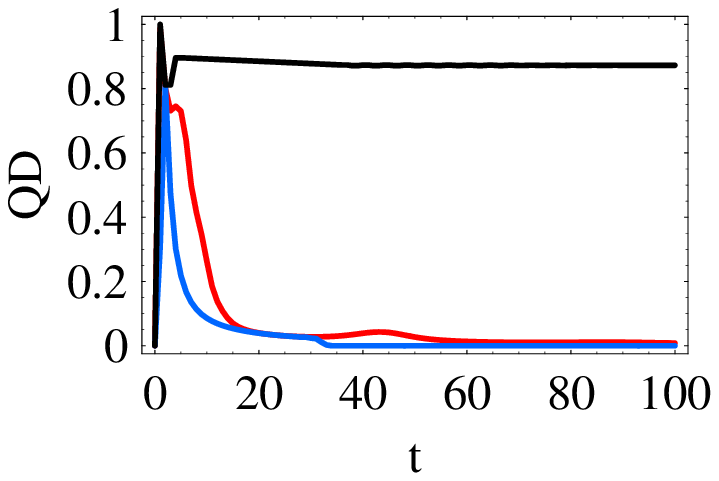}
   \includegraphics[width=9cm]{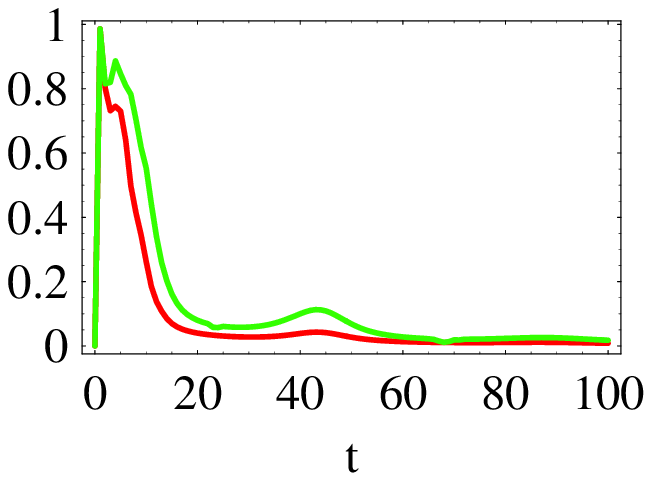}
   \caption{(Color online) (a) The MID, (b) QD for a QW on a line after~$t=100$ steps with
   perfect (in black), non-Markovian dephasing coin (in red) and Markovian dephasing coin (in blue), respectively.
   Here we choose the parameters $g_0=1$, $\eta=0.01$ and $\eta=10$ respectively for calculations.
   (c) The comparison of MID in red and QD in green for a QW on a line with the same parameters above.}
   \label{MID}
\end{figure}

We numerically calculate the MID and QD for a QW with perfect,
Markovian and non-Markovian dephasing coins in the short time
regime. The quantum correlation between quantum walker and his coin
presented by the MID and QD is oscillating for the first several
steps ($\lesssim 100$) shown in Fig.~\ref{MID}(a) and (b). Without
decoherence there is no difference between MID and QD. Memory
effects caused by the non-Markovianity of the environment deviate
the MID and QD from monotonically behavior as that during the
Markovian process. Both MID and QD decay steeply and interrupted by
revivals. The period of oscillating is as same as that of
corresponding position variance, which proves that quantum
correlation between quantum walker and his coin results in the
behavior of the QW. Either of the MID and QD tends to zero with time
increasing. Whereas for ideal QWs, both MID and QD do not decrease
with time as shown in~\cite{SBC1,SBC2}. From the numerical results
one can see the quantum correlation in the case of Markovian
decoherence is the lower bound of that with Non-Markovian
decoherence, which explains the reason that the position variance of
a RW distribution provides the lower bound of that of QW with
non-Markovian dephasing coin. In comparison of MID and QD shown in
Fig.~\ref{MID}(c), both of them can show the quantum correlation
between quantum walker and his coin changing with time and MID
oscillates with bigger amplitude than QD, which makes that the MID
might be better signature for observing in this case.

\section{A QW on a line with a non-Markovian dephasing coin in the long time limit}

In the above sections we show the behavior of QW on a line with a
non-Markovian dephasing coin results in the quantum correlation
between quantum walker and his coin, which is presented by the MID
and QD. We observe that both MID and QD decrease with time and tend
to zero in the long time limit. That means the quantum correlation
will disappear for long enough time. In this section, we will show
in the long time regime, the behavior of a non-Markovian decoherent
QW.

The probability for the walker to reach a point $x$ at time $t$ is
\begin{align}
P(x;t)&=\text{Tr}\{\ket{x}\bra{x}\rho(t)\}\nonumber\\
&=\frac{1}{(2\pi)^2}\int_{-\pi}^\pi \text{d}k\int_{-\pi}^\pi
\text{d}k'
e^{-i(k-k')x}\text{Tr}\{\mathcal{L}^t_{kk'}\ket{\Phi_0}\bra{\Phi_0}\}.
\end{align}
Thus we can calculate the moments of this distribution:
\begin{align}
\langle x^m \rangle&=\sum_x x^m P(x;t)
\nonumber\\&=\frac{(-1)^m}{(2\pi)^2}\int_{-\pi}^\pi
\text{d}k\int_{-\pi}^\pi \text{d}k'
\delta^{(m)}(k-k')\text{Tr}\{\mathcal{L}^t_{kk'}\ket{\Phi_0}\bra{\Phi_0}\}.
\end{align}
We can then invert the order of operations and do the $x$ sum first.
This sum can exactly be carried out in terms of derivatives of the
$\delta$ function:
\begin{equation}
\frac{1}{2\pi}\sum_x x^m e^{-ix(k-k')}=(-i)^m\delta^{(m)}(k-k').
\end{equation}
Inserting this result back into the expression for $\langle
x\rangle$ we get the first moment
\begin{align}
\langle x\rangle&=\frac{-i}{2\pi}\int_{-\pi}^{\pi} \text{d}k
\frac{d}{\text{d}k}\text{Tr}\{\mathcal{L}^t_{k}\ket{\Phi_0}\bra{\Phi_0}\}\nonumber\\
&=-\frac{1}{2\pi}\int_{-\pi}^{\pi}\text{d}k\sum_{j=1}^t\text{Tr}\{\sigma_z
\mathcal{L}^j_{k}\ket{\Phi_0}\bra{\Phi_0}\}. \label{eq:first}
\end{align}
We can carry out a similar integration by parts to get the second
moment:
\begin{equation}
\langle
x^2\rangle=-\frac{1}{2\pi}\int_{-\pi}^{\pi}\text{d}k\left\{\sum_{j=1}^t\sum_{j'=1}^j
\text{Tr}\left\{\sigma_z\mathcal{L}^{j-j'}_{k}(\sigma_z\mathcal{L}_{k}^{j'}\ket{\Phi_0}\bra{\Phi_0})\right\}+\sum_{j=1}^t\sum_{j'=1}^{j-1}
\text{Tr}\left\{\sigma_z\mathcal{L}^{j-j'}_{kj}\left[(\mathcal{L}_{k}^{j'}\ket{\Phi_0}\bra{\Phi_0})\sigma_z\right]\right\}\right\}.
\end{equation}

Because $\mathcal{L}_k$ is a linear transformation we can represent
it as a matrix acting on the space of $2\times 2$ operators. We
choose the representation as
\begin{equation}
\hat{O}=r_1\one+r_2\sigma_x+r_3\sigma_y+r_4\sigma_z.
\end{equation}
The action of $\mathcal{L}_{k}$ on $\hat{O}$ is given by the matrix
\begin{align}
\mathcal{L}_{k}\hat{O}
        =\begin{pmatrix}
           1 & 0 & 0 & 0\\
           0 & \sin2k \text{Im}\left[\kappa(t)\right] & \sin2k\text{Re}\left[\kappa(t)\right] & \cos2k\\
           0 & -\cos2k \text{Im}\left[\kappa(t)\right] & -\cos2k \text{Re}\left[\kappa(t)\right] &
           \sin2k\\
           0 & \text{Re}\left[\kappa(t)\right] & -\text{Im}\left[\kappa(t)\right]&
           0
        \end{pmatrix}\begin{pmatrix}
           r_1 \\r_2\\ r_3 \\r_4
        \end{pmatrix}
\end{align}
Because
$r_1=\text{Tr}\left(\hat{O}\right)=\text{Tr}\left(\mathcal{L}_k\hat{O}\right)$.
The only nontrivial dynamics results from the $3\times 3$ submatrix
\begin{align}
M_k=\begin{pmatrix}
           0 & 0 & 0\\
           \sin2k \text{Im}\left[\kappa(t)\right] & \sin2k\text{Re}\left[\kappa(t)\right] & \cos2k\\
           -\cos2k \text{Im}\left[\kappa(t)\right] & -\cos2k \text{Re}\left[\kappa(t)\right] &\sin2k\\
           \text{Re}\left[\kappa(t)\right] & -\text{Im}\left[\kappa(t)\right]&0
        \end{pmatrix}
\end{align}

We also need to know the effects of left and right multiplying by
$\sigma_z$. These are given by the two matrices
\begin{align}
Z_L=\begin{pmatrix}
           0 & 0 & 0 & 1\\
           0 & 0 & i & 0\\
           0 & -i & 0 & 0\\
           1 & 0 & 0 & 0
        \end{pmatrix},Z_R=\begin{pmatrix}
           0 & 0 & 0 & 1\\
           0 & 0 & -i & 0\\
           0 & i & 0 & 0\\
           1 & 0 & 0 & 0
    \end{pmatrix}.
\end{align}

Let us take these expressions and apply them to Eq.~(\ref{eq:first})
for the first moment. In the integrand, the initial density matrix
for the coin is multiplied $j$ times by $\mathcal{L}_k$, then left
multiplied by $\sigma_z$ and finally the trace is taken. Given the
above expression for $Z_L$, we see that this is the same as
multiplying the $3$-vector $\left(r_2,r_3,r_4\right)$ $j$ times by
$M_k$ and then keeping only the $r_4$ component of the result. This
gives us the new expression for the first moment
\begin{align}
\langle
x\rangle&=-\frac{1}{2\pi}\int_{-\pi}^{\pi}\text{d}k\begin{pmatrix}
0&0&1 \end{pmatrix}\left(\sum_{j=1}^t M_k^j\right)\begin{pmatrix}
r_2\\r_3\\r_4
\end{pmatrix}\\&=-\frac{1}{2\pi}\int_{-\pi}^{\pi}\text{d}k\begin{pmatrix}
0&0&1
\end{pmatrix}\left[\left(1-M_k\right)^{-1}\left(M_k-M_k^{t+1}\right)\right]\begin{pmatrix}
r_2\\r_3\\r_4
\end{pmatrix}.
\end{align}
The eignevalues of $M_k$ satisfy $0<|\lambda|<1$ and then in the
long time limit, therefore $M_k^{t+1}\rightarrow 0$ and the first
moment becomes approximately
\begin{align}
\langle x\rangle &\approx
-\frac{1}{2\pi}\int_{-\pi}^{\pi}\text{d}k\begin{pmatrix} 0&0&1
\end{pmatrix}\left[\left(1-M_k\right)^{-1}M_k\right]\begin{pmatrix}
r_2\\r_3\\r_4
\end{pmatrix}\\&=\frac{r_2\text{Re}\left[\kappa(t)\right]-r_3\text{Im}\left[\kappa(t)\right]+r_4|\kappa(t)|^2}{-1+|\kappa(t)|^2}.
\end{align}

In the long time limit, the second moment can also be calculated as
\begin{align}
\langle
x^2\rangle&=t-\frac{1}{2\pi}\int_{-\pi}^{\pi}\text{d}k\begin{pmatrix}1&0&0&0
\end{pmatrix}\left[Z_L\sum_{j=1}^t\sum_{j'=1}^{j-1}\mathcal{L}_k^{j-j'}\left(Z_L+Z_R\right)\mathcal{L}_k^{j'}\right]\begin{pmatrix}
1\\r_2\\r_3\\r_4 \end{pmatrix}\nonumber\\
&=t-\sum_{j=1}^t\sum_{j'=1}^{j-1}\frac{1}{2\pi}\int_{-\pi}^{\pi}\text{d}k\begin{pmatrix}
0&0&1 \end{pmatrix}M_k^{j-j'}\begin{pmatrix} 0\\0\\2
\end{pmatrix}\nonumber\\
&=t-\frac{1}{2\pi}\int_{-\pi}^{\pi}\text{d}k\begin{pmatrix} 0&0&1
\end{pmatrix}\left[t-\left(1-M_k\right)^{-1}M_k+\left(1-M_k\right)^{-1}M_k^t\right]\left(1-M_k\right)^{-1}M_k\begin{pmatrix} 0\\0\\2
\end{pmatrix}\nonumber\\
&=t\left(1+\frac{2|\kappa(t)|^2}{-1+|\kappa(t)|^2}\right)+\frac{2|\kappa(t)|^4+5|\kappa(t)|^2}{\left(-1+|\kappa(t)|^2\right)^2},
\end{align}
which is independent on the initial coin state.

The position variance is obtained as
\begin{align}
\text{var}=\langle x^2\rangle-\langle
x\rangle^2=t\left(1+\frac{2|\kappa(t)|^2}{-1+|\kappa(t)|^2}\right)+\frac{2|\kappa(t)|^4+5|\kappa(t)|^2}{\left(-1+|\kappa(t)|^2\right)^2}-\frac{r_2\text{Re}\left[\kappa(t)\right]-r_3\text{Im}\left[\kappa(t)\right]+r_4|\kappa(t)|^2}{-1+|\kappa(t)|^2}.
\end{align}
In the long time regime, $\kappa(t)$ (\ref{eq:kappa}) tends to zero.
Then the position variance is linear dependent on time $t$. Thus in
the case of the non-Markovian dephasing coin the variance grows
linearly with time and a QW with a non-Markovian dephasing coin
shows a diffusive spread in the long time limit.

\begin{figure}
   \includegraphics[width=9cm]{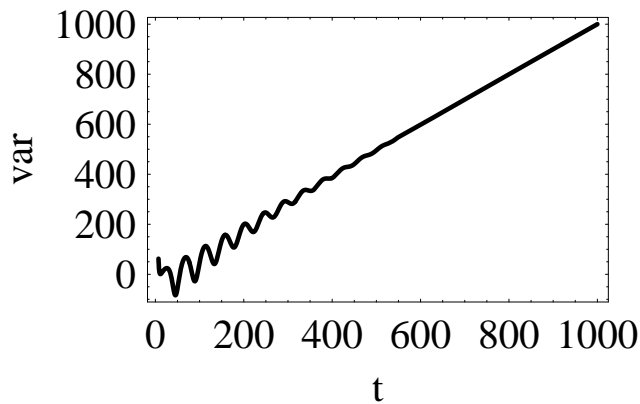}
   \caption{The analytical result on the variance of position distribution for a
   QW on a line with a non-Markovian dephasing coin in the long time regime.
   Here we choose the parameters $g_0=1$ and $\eta=0.01$ for calculations.}
   \label{ana}
\end{figure}

In the long time limit, we can obtain the analytical results of the
position variance of a non-Markovian decoherent QW. In
Fig.~\ref{ana}, we can see that for the first several steps, the
numerical and analytical results agree with each other. Whereas for
long time, the variance becomes linear dependent on time $t$ showing
a diffusive spread, just as in the classical case, though the rate
of growth will greater than for the RW.

\section{Conclusion}
In this paper, we presented how non-Markovian decoherence can
influence the evolution of a quantum walker on a line and showed a
different behavior compared to the ballistic and diffusive spreads.
We analyze the dynamics and quantum correlation of quantum walker
and his coin. We use the MID and QD as measures for quantum
correlations. The QD would be a valuable measure to use and the MID
also suffices to show that pure quantum correlation induced by
having non-Markovian dephasing on the coin state. The memory effect
caused by the information flowing back to the system during a
non-Markovian process decreases the quantum correlation quickly
which is later interrupted by revivals. In the long time limit, the
quantum correlation disappears with $\kappa(t)$ tending to zero,
which results in a transition from a QW with a ballistic scaling to
a RW with a diffusive scaling. We draw the conclusions as
followings: in the short time the behavior of QW with non-Markovian
dephasing coin oscillates between a ballistic and diffusive
spreading behavior, and the quantum correlation between the walker
and coin collapses and revivals due to the memory effect. Whereas in
the long time limit, QW with non-Markovian decoherence does not make
much difference from that with Markovian decoherence. Thus our
analysis is quite valuable in that we characterize this QW on a line
with non-Markovian dephasing coin carefully and devise appropriate,
meaningful QD and MID as measures to study quantum correlation
dynamics for this system.

\begin{acknowledgments}
We would like to thank Barry C. Sanders for useful conversations.
This work has been supported by the National Natural Science
Foundation of China under Grant Nos 10974192, 11004029 and 11174052,
the Natural Science Foundation of Jiangsu Province under Grant No
BK2010422, the Ph.D.\ Program of the Ministry of Education of China,
the Excellent Young Teachers Program of Southeast University and the
National Basic Research Development Program of China (973 Program)
under Grant No 2011CB921203.
\end{acknowledgments}

\end{document}